# Void-Defect Induced Magnetism and Structure Change of Carbon Materials - I : Graphene Nano Ribbon

Norio Ota and Laszlo Nemes*

Graduate School of Pure and Applied Sciences, University of Tsukuba, *1-1-1 Tenoudai Tsukuba-city 305-8571, Japan*,
*Research Center for Natural Sciences, Ötvös Lóránd Research Network, *Budapest 1519, Hungary*

Void-defect is a possible origin of ferromagnetic-like feature of pure carbon material. Applying density functional theory to void-defect induced graphene-nano-ribbon (GNR), a detailed relationship between multiple-spin-state and structure change was studied. An equilateral triangle of an initial void having six electrons is distorted to isosceles triangle by re-bonding carbon atoms. Among possible spin-states of $S_z$=0/2, 2/2, 4/2 and 6/2, the most stable state was $S_z$=2/2. The case of $S_z$=4/2 is remarkable that initial flat ribbon turned to three dimensionally curled one having highly polarized spin configuration at ribbon edges. Total energy of $S_z$=4/2 was very close to that of $S_z$=2/2, which suggests coexistence of flat ribbon and curled ribbon. As a model of three dimensional graphite, bi-layered AB stacked GNR was analyzed for cases of different void position of α-site and β-site. Spin distribution was limited to the surface layer, nothing to the back layer. Distorted void triangle show 60degree clockwise rotation from α- to β-site, which was consistent with experimental observation using the scanning tunneling microscope. This study revealed that void-defect in GNR induces unusual polarized spin state, different with usual ferromagnetic one.
[ To be published on Journal of the Magnetics Society of Japan (2021), e-mail to Norio Ota: n-otajitaku@nifty.com   ]

## 1. Introduction

These ten years, it was reported that some carbon base materials show room temperature ferromagnetic like hysteresis[1)-6)]. They are graphite and graphene like materials. Such a light-weight ferromagnetic like materials will be useful for many applications. However, such magnetic ordering could not be thoroughly understood. Possible explanations are the presence of impurities[7)], edge irregularities[8)-10)] or void-defects[11)-16)]. Here, we like to focus on void-defect. In experiments, creation of void-defect was done by high energy particle irradiation on graphite[11)-13)]. Void-defect was observed by the scanning tunneling microscope[14)-16)]. Unfortunately, there are little information on magnetic behaviors and spin distributions in atomic scale. Theoretical calculations predicted the importance of the atomic structure change[17)-18)]. However, there are little explanation on a detailed relationship between the multiple-spin-state and the structure-change. Here, we try detailed explanation by the density functional theory (DFT) [19)-20)]. DFT gives the stable quantum state basically at zero temperature. In this study, we like focus on fundamental property of void-defect induced magnetism and structure change of graphene nano ribbon (GNR). Such analysis will be useful for future advanced study on room temperature magnetism and asking Curie-temperature.

There are six unpaired electrons around one void, which suggests a capability of multiple spin-state as like $S_z$=6/2, 4/2, 2/2, and 0/2. We need detailed spin dependent calculation to find the most stable spin state accompanying with the structure change Also, we will calculate the three-dimensional graphite using the model of AB-stacked bi-layer GNR. Results will be compared with experimental observation by the scanning tunneling microscope.

## 2. Calculation Method and Model GNR

We require total energy, optimized molecular structure and the spin density on a given spin state of $S_z$. DFT based generalized gradient approximation (GGA-PBEPBE) [21)] was applied utilizing Gaussian09 package[22)] with an atomic orbital 6-31Gd basis set[23)]. Total charge is set to be completely zero. Inside of a super-cell, three dimensional DFT was applied. As illustrated in Fig. 1, one dimensional periodic boundary condition was applied to realize an unlimited length GNR. Calculation is repeated until to meet the convergence criteria on the root mean square density matrix less than 10$^{-8}$ within 128 cycles.

An initial calculation model of GNR with single void-defect is shown in left of Fig. 1. Super-cell was [$C_{79}H_{10}$], where upper and lower zigzag edge carbons were all hydrogenated to avoid radical carbon edge complicated situation. Ribbon width was 1.780nm and one-dimensional super-cell length was 1.241nm repeating to red-line direction. A center positioned carbon was removed to make a void-defect. Around this void, there are three carbons, which make an equilateral triangle with length of 0.248nm. These three radical carbons bring six electrons. To investigate magnetic characteristics, we should consider these six electrons interaction, which mean to study a detailed multiple-spin-state analysis.

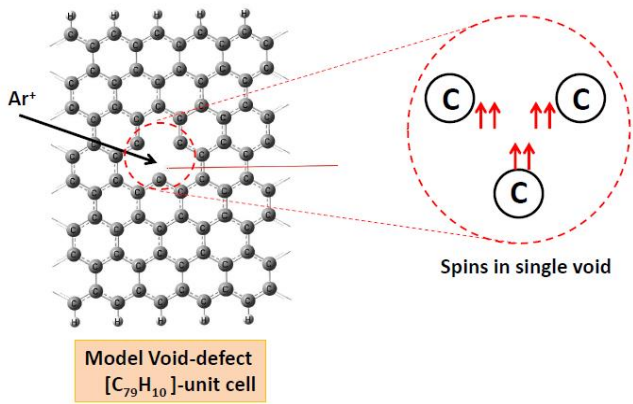

**Fig. 1** Initial void-defect is created on GNR. Six spins in a void carry the multiple spin state. Unit cell is [$C_{79}H_{10}$]. DFT calculation is done three dimensionally inside of unit cell, which is repeated one dimensionally toward a red line direction.

.

### 3. Multiple Spin State

For six electrons in a void-defect, there are four capable spin states of $S_z$=0/2, 2/2, 4/2 and 6/2. However, we could not get any converged calculation on singlet spin state of $S_z$=0/2. The reason will be explained as illustrated on right of Fig. 1 that one radical carbon holds two spins, which should be parallel to avoid unlimited large coulomb energy due to Hund's rule[24], in case of one central attractive force by the same carbon atom. Three pairs of parallel spins enable the multiple-spin-state of $S_z$=6/2, 4/2, or 2/2. For every spin state, stable atomic structures are classified to two types as Type-A and -B as shown in Fig. 2 as (b) and (c). It should be noted that initial equilateral triangle (a) turned to an isosceles one of (b) and (c). Isosceles triangle of Type-A was perpendicular to a ribbon axis of a red line, whereas Type-B tilted. Such distortion originates from the quantum mechanical re-bonding as discussed by Yazyev and Helm[17]. Detailed results were shown in Fig. 3 and Table 1. Initial non-deformed ribbon energy was defined to be zero. In Type-A, the most stable spin state was $S_z$=2/2. Energy level was reduced to -12.31kcal/mol., where triangle has an angle of 49degree as shown in (b). Type-B show remarkable energy reduction. In case of $S_z$=2/2 of Type-B, energy became -15.62kcal/mol, which is most stable one with a sharper angle of 44degree in (c). Distance between carbon-atom "a" and "c" was shortened to 0.192nm from original 0.248nm. This suggests a capable origin of structure change that original pi-electron conjugated bond is partially modified to double-bond among hexagon carbon networks.

Sum of spin density around initial equilateral void was 1.49μ$_B$, which coincides well with calculation using the SIESTA code by Yazyev and Helm[17]. Spin densities of Type-A ($S_z$=2/2) and B ($S_z$=2/2) were illustrated in Fig. 4 at a surface of electron density of 1e/nm$^3$. In both cases, we can see a large up-spin cloud (red) at a void triangle's apex carbon site. We can see fruit pear like up-spin cloud with sum of 1.37μ$_B$, which was 0.12μ$_B$ smaller than initial one. Inside of GNR, there appears up (red) and down (blue) spins alternatively[9-10].

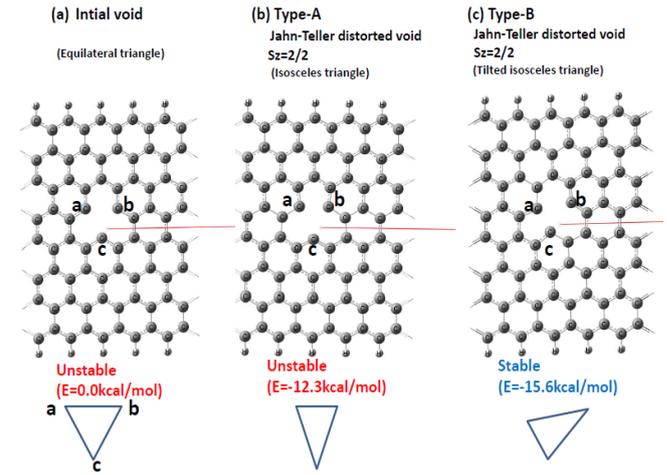

**Fig. 2** Initial void makes equilateral triangle. Type-A has an isosceles triangle perpendicular to GNR, whereas Type-B tilted.

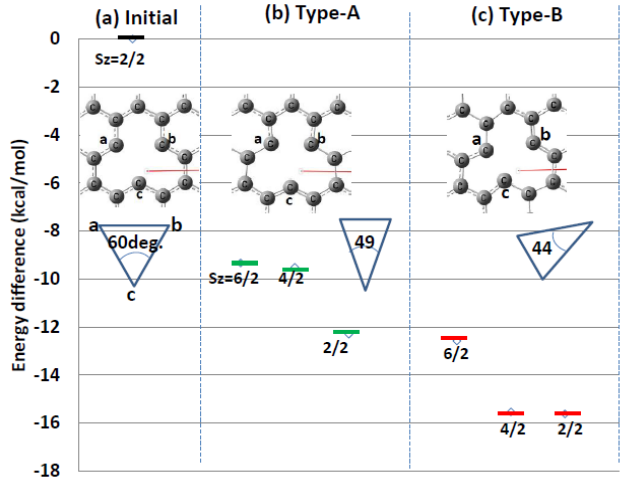

**Fig. 3** Energy difference and void structure change. Type-A show distorted isosceles void triangle of 49degree perpendicular to ribbon as shown in (b). Type-B shows a tilted triangle with 44degree. In every type, triplet spin-state of $S_z$=2/2 is most stable.

**Table 1** Calculated result for mono layer GNR.

| Distorted type | none | Type A | | | Type B | | |
|---|---|---|---|---|---|---|---|
| Case number | A0 | A1 | A2 | A3 | B1 | B2 | B3 |
| Given Sz | 2/2 | 6/2 | 4/2 | 2/2 | 6/2 | 4/2 | 2/2 |
| Energy difference (kcal/mol/unit cell) | 0 | -9.33 | -9.51 | -12.31 | -12.58 | -15.54 | -15.62 |
| Ribbon configuration | Flat | Flat | Flat | Flat | Flat | Curled | Flat |
| Triangle distance ab (Å) | 2.48 | 2.15 | 2.11 | 2.14 | 2.59 | 2.61 | 2.59 |
| bc | 2.48 | 2.6 | 2.59 | 2.6 | 2.61 | 2.64 | 2.61 |
| ca | 2.48 | 2.6 | 2.59 | 2.6 | 1.96 | 1.82 | 1.92 |
| Smallest angle (deg.) | 60 | 48.8 | 48.1 | 48.6 | 44.3 | 40.6 | 43.5 |
| Mulliken charge (e) a | 0.21 | 0.25 | 0.25 | 0.25 | 0.25 | 0.22 | 0.25 |
| b | 0.21 | 0.25 | 0.25 | 0.25 | 0 | 0 | -0.01 |
| c | 0.22 | 0.05 | 0.03 | 0.04 | 0.27 | 0.23 | 0.27 |
| Spin density ($\mu_B$) a | -0.08 | 0.24 | 0.03 | 0.21 | 0.19 | 0.01 | 0.15 |
| b | 0.77 | 0.24 | 0.03 | 0.21 | 1.09 | 0.79 | 1.06 |
| c | 0.97 | 1.15 | 0.81 | 1.12 | 0.34 | -0.17 | 0.28 |
| a+b+c | 1.66 | 1.63 | 0.87 | 1.54 | 1.62 | 0.63 | 1.49 |
| Ribbon magnetism | Ferrimag. | Ferrimag. | Ferromag. | Ferrimag. | Ferrimag. | Ferromag. | Ferrimag. |

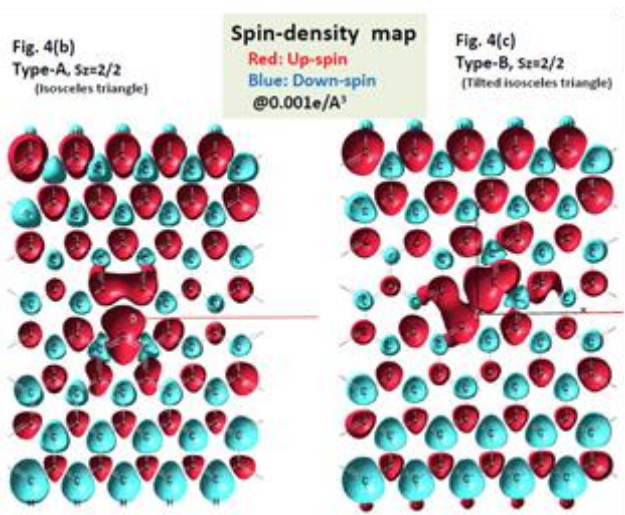

**Fig. 4** Irregular up-spin cloud (by red) was observed at centered void triangle for both Type-A and -B. Inside of ribbon, there appear alternate up-spin (red) and down-spin (blue).

## 4. Curled ribbon

Amazing result was obtained in case of $S_z$=4/2 of Type-B as illustrated on Fig. 5(d), where appears a curled ribbon. While in case of $S_z$=2/2, ribbon is flat in (c). Void triangle of $S_z$=4/2 was 40.6degree comparing with 43.5 degree of $S_z$=2/2. The spin distribution is ferromagnetic like for $S_z$ =4/2 as illustrated in Fig. 6(d). Both of upper and lower ribbon edges carry up-spin clouds. While in case of $S_z$=2/2 of Fig. 6(c), we can see anti-ferromagnetic like feature. It should be noted that energy of $S_z$=2/2 was -15.6kcal/mol., whereas $S_z$=4/2 was -15.5kcal/mol. They are only 0.1kcal/mol. difference. Such close energy suggests the coexistence both of flat ribbon and curled ribbon. Spin configuration of Fig 6(d) looks ferromagnetic like arrangement. Edge carbons are strongly magnetized by up-spins. However, it should be noted that each up-spins are isolated with neighbor edge spins. There is no exchange interaction between them. On the contrary, as studied in our previous paper on FeO-modified GNR[25], edge carbons were highly polarized by up-spins, and also coupled with neighbor edge spins by super-exchange interaction. It was usual ferromagnetic coupling. By such comparison, void induced GNR show unusual strongly polarized spin configuration, different with usual ferromagnetic one.

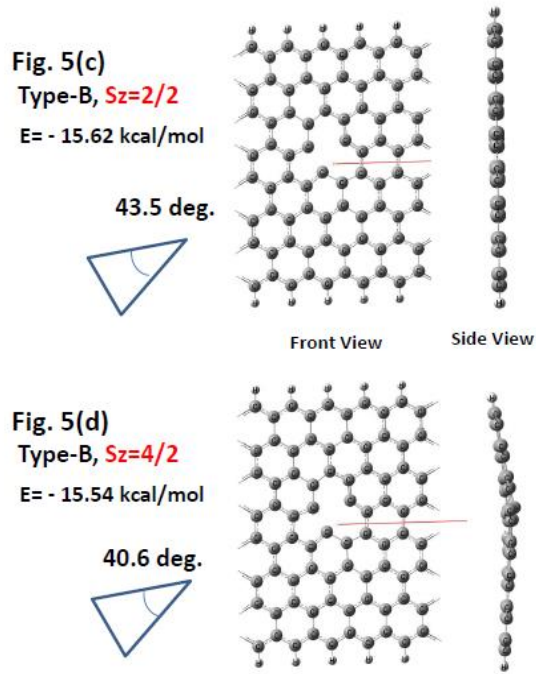

**Fig. 5** In case of $S_z=4/2$ of Type-B, ribbon is curled as shown in (d), whereas for $S_z=2/2$ ribbon is flat (c).

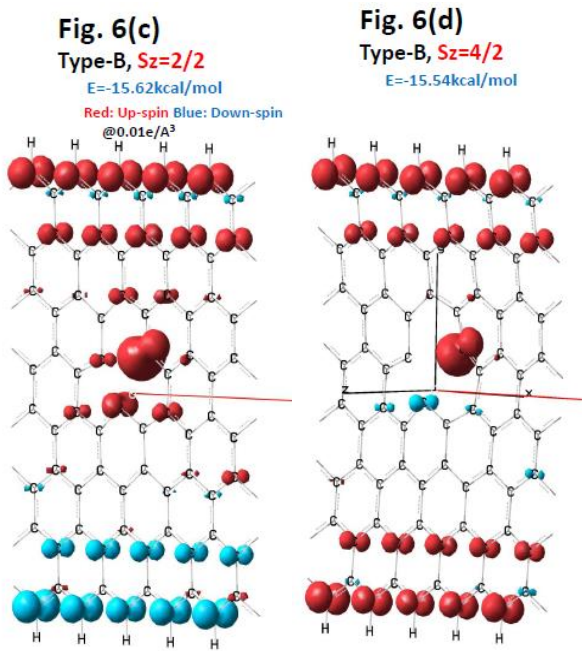

**Fig. 6** In case of $S_z=2/2$ for Type-B, spin configuration is ferrimagnetic-like as shown in (c). Whereas in case of $S_z=4/2$, it is ferromagnetic-like (d).

## 5. Bi-layer GNR

To simulate three-dimensional graphite, bi-layered AB stacked GNR was analyzed. In Fig. 7, yellow marked atoms are surface (first) layer atoms including single void, whereas gray atoms are back (second) layer without any void. There are two types of void positions, which one is α-site void positioned over one carbon of the second layer (back side), whereas β-site positioned over inter-atom space. Unit cell is [$C_{79}H_{10}$-$C_{80}H_{10}$]. Distance between two layers was calculated to be from 0.409 to 0.413nm depending on void situation as shown in Table 2. We can see deformed void triangle as classified by Type-A and Type–B as shown in Fig. 8. In every Type, the most stable spin state was $S_z=2/2$. Total energy of Type-B was lower (stable) than that of Type-A. Distorted triangle of Type-B shows small angle of 42degree compared with 53degree of Type-A. It should be noted that, in case of Type-B, total energy is almost similar for α-site (-14.59kcal/mol.) and β-site (-14.69kcal/mol.). We can expect coexistence of both sites. Detailed results are summarized in Table 2. In Fig. 9, spin density of AB-stacked bi-layer GNR was illustrated. Spin density appeared only to the surface (first) layer. Back (second) layer shows no spin. Summed spin-density around one void was 1.37μB.

## 6. Comparison with experiments

### 6.1 Observation of void-defect

In 1998, Kerry et al.[14] observed void-defect on the surface of graphite by the scanning tunneling microscope. They observed triangle shape bright spot, which suggested large excess electrons inside of triangle void. Our calculation is consistent with such observation.

### 6.2 Observation of rotational symmetry of void defect

Notable result is the clockwise angle of a void triangle defined in the figure on bottom of Table 2. The clockwise angle θ of α-site void triangle was 300degree, whereas β-site 240degree. There appear 60degrees rotation from β-site to α-site. Such calculation is consistent with observation on graphite surface done by Ziatdinov et al.[16] using the scanning tunneling microscope.

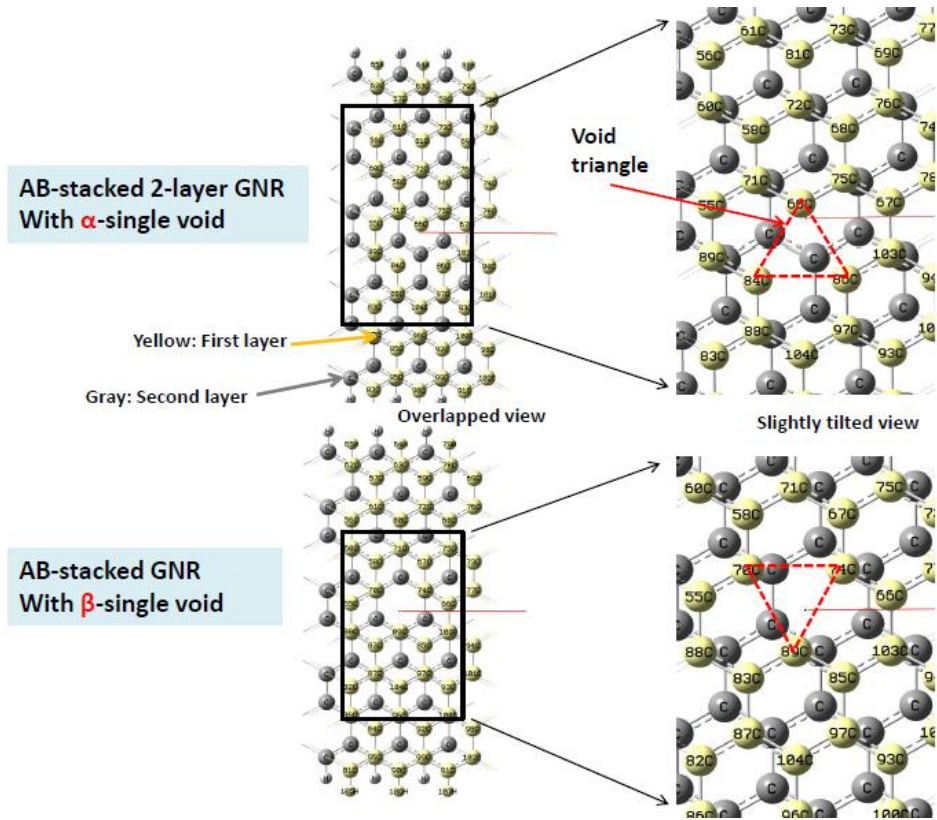

**Fig. 7** Void-defect was created on AB-stacked bi-layer GNR with a void at α-site (top panel) and β-site (bottom). Yellow atoms are surface layer carbon, whereas gray atoms back layer one.

**Table 2** Calculated results of bi-layer GNR with single void. Cases are classified for Type-A or Type-B, and for void positions of α-site or β-site.

| Case number | Bi-initial | Alpha-A | Alpha-B | Beta-A | Beta-B |
|---|---|---|---|---|---|
| Void site | β | α | α | β | β |
| Distorted type | none | Type A | Type B | Type A | Type B |
| Given Sz | 2/2 | 2/2 | 2/2 | 2/2 | 2/2 |
| Stacked distance d (Å) | 4.1 | 4.09 | 4.11 | 4.13 | 4.11 |
| Energy difference (kcal/mol/unit cell) | 0 | −13.33 | −14.59 | −12.61 | −14.69 |
| Ribbon configuration | Flat | Flat | Flat | Flat | Flat |
| Triangle distance ab (Å) | 2.45 | 2.63 | 2.64 | 2.34 | 2.53 |
| bc | 2.49 | 2.34 | 2.53 | 2.63 | 2.64 |
| ca | 2.49 | 2.63 | 1.84 | 2.63 | 1.84 |
| Smallest angle (deg.) | 60 | 52.8 | 41.7 | 52.8 | 41.7 |
| Mulliken charge (e) a | 0.22 | 0.04 | 0.21 | 0.21 | 0.24 |
| b | 0.22 | 0.21 | −0.01 | 0.21 | −0.01 |
| c | 0.17 | 0.21 | 0.24 | 0.04 | 0.21 |
| Spin density ($\mu_B$) a | 0.33 | 0.9 | 0.04 | 0.25 | 0.25 |
| b | 0.34 | 0.25 | 1.08 | 0.25 | 1.08 |
| c | 0.84 | 0.25 | 0.25 | 0.9 | 0.04 |
| Spin density (a+b+c) | 1.51 | 1.4 | 1.37 | 1.4 | 1.37 |
| Ribbon magnetism | Ferrimag. | Ferrimag. | Ferrimag. | Ferrimag. | Ferrimag. |
| Triangle vacancy tilt angle against GNR ($\theta$) | 0 deg. | 180 | 300 | 0 | 240 |

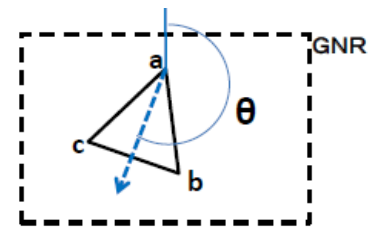

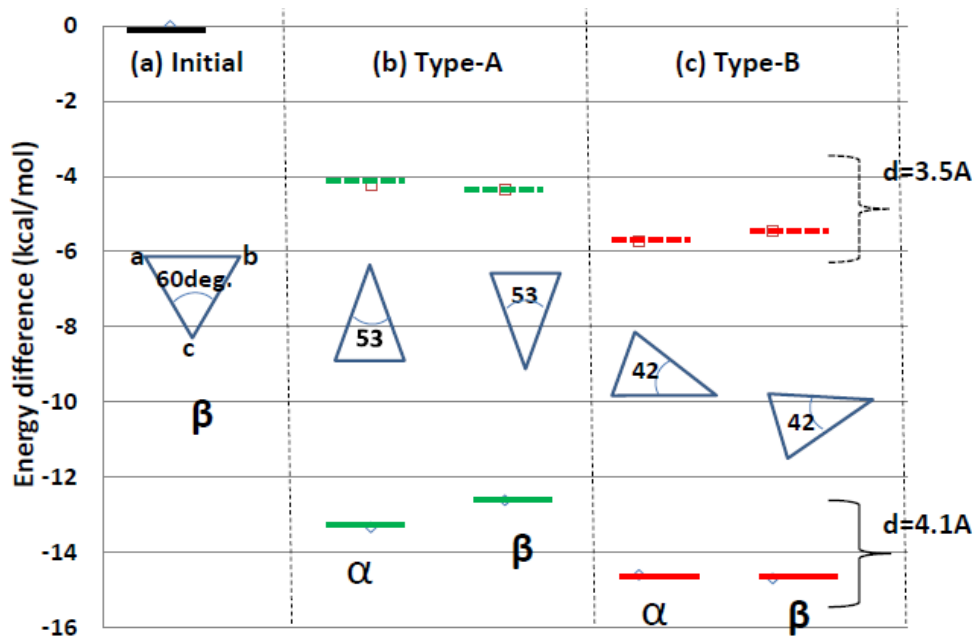

**Fig. 8** Energy and void triangle are compared in AB-stacked bi-layer GNR.

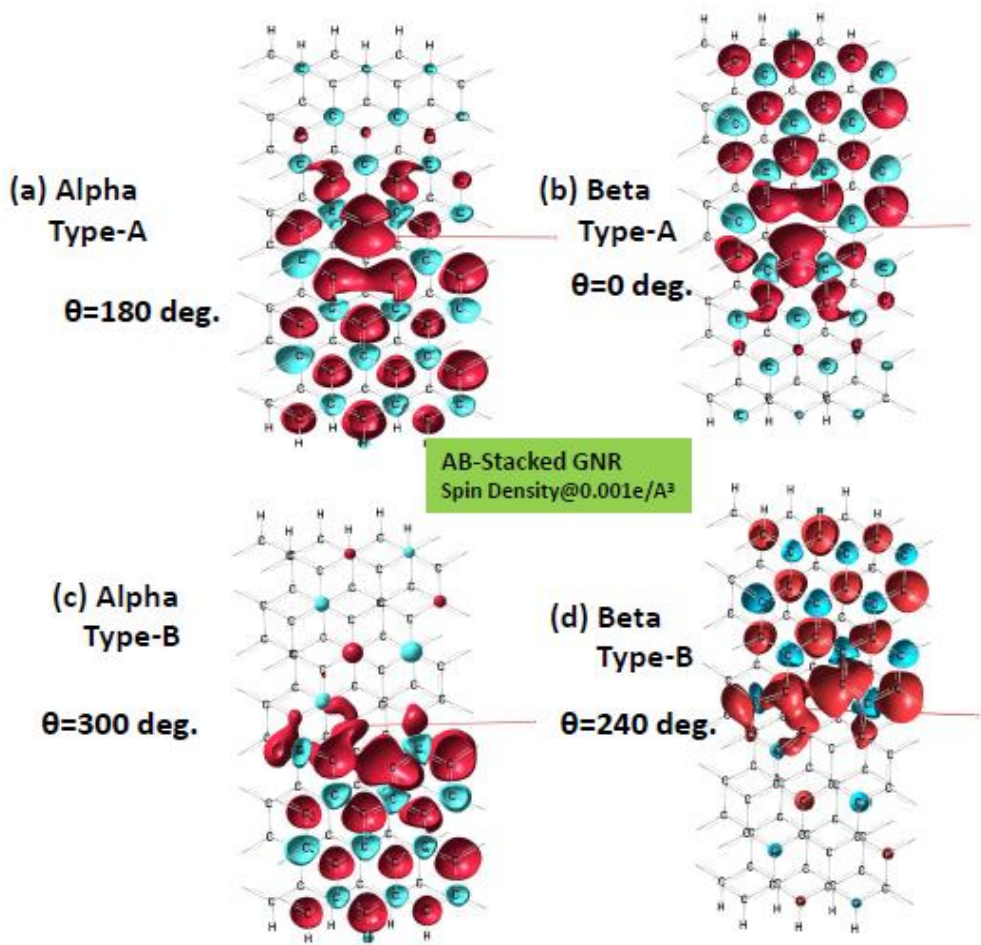

**Fig.9** Spin density configuration of AB-stacked bi-layer GNR, where red cloud shows up-spin and blue one down-spin. Spin appears only to the surface layer. Back layer shows no spin.

# 7. Conclusion

Applying density functional theory to a void-defect in graphene-nano-ribbon (GNR), a relationship between multiple-spin-state and structure change was studied. An equilateral triangle of initial void was distorted to isosceles triangle due to re-bonding of excess six electrons in a void. Such calculation is consistent with actual observation by the scanning tunneling microscope. Six electrons enable the multiple spin-state of $S_z$=6/2, 4/2 and 2/2 due to the Hund's rule. The most stable spin state was $S_z$=2/2 showing flat ribbon structure. Amazing result was obtained for $S_z$=4/2, where ribbon was three dimensionally curled, and show ferromagnetic like spin distribution on both upper and lower edges of the ribbon. Energy of $S_z$=2/2 and $S_z$=4/2 were close, almost same. This suggests coexistence of flat and curled ribbon. As a model of three-dimensional graphite, bi-layered AB stacked GNR was analyzed for different void positions of α-site and β-site. Both sites show similar energy. Void triangle presents 60 degrees clockwise rotation from β-site to α-site. It should be noted that such void rotation coincides well with actual observation. This study revealed that void-defect in GNR induces unusual highly polarized spin state, different with usual ferromagnetic one.

# References

1) P. Esquinazi, D. Spemann, R. Hohne, A. Setzer, K. Han, and T. Butz: *Phys. Rev. Lett.*, **91**, 227201 (2003).
2) K. Kamishima, T. Noda, F. Kadonome, K. Kakizaki and N. Hiratsuka: *J. of Mag. and Magn. Mat.*, **310**, e346 (2007).
3) T. Saito, D. Nishio-Hamane, S. Yoshii, and T. Nojima: *Appl. Phys. Lett.*, **98**, 052506 (2011).
4) Y. Wang, Y. Huang, Y. Song, X. Zhang, Y. Ma, J. Liang and Y. Chen: *Nano Letters*, **9**, 220 (2009).
5) J. Cervenka, M. Katsnelson and C. Flipse: *Nature Phys.*, **5**, 840 (2009), (https://doi.org/10.1038/nphys1399) .
6) H. Ohldag, P. Esquinazi, E. Arenholz, D. Spemann, M. Rothermal, A. Setzer, and T.Butz: *New Journal of Physics*, **12**, 123012 (2010).
7) J. Coey, M. Venkatesan, C. Fitzgerald, A. Douvalis and I. Sanders: *Nature*, **420**, 156 (2002).
8) K. Kusakabe and M. Maruyama: *Phys. Rev. B*, **67**, 092406 (2003).
9) N. Ota, N. Gorjizadeh and Y. Kawazoe: *J. Magn. Soc. Jpn.*, **36**, 36 (2012).
10) N. Ota: *J. Magn. Soc. Jpn.*, **37**, 175 (2013).
11) P. Lehtinen, A. Foster, Y. Ma, A. Krasheninnikov, and R. Nieminen: *Phys. Rev. Lett.*, **93**, 187202 (2004).
12) P.Ruffieux, O. Groning, P. Schwaller, L. Schlapbach, and P. Groning: *Phys. Rev. Lett.*, **84**, 4910 (2000).
13) A. Hashimoto, K. Suenaga, T. Sugai, H.Shinohara, and S. Iijima: *Nature (London)*, **430**, 870 (2004).
14) K.Kelly and N.Hales: *Surface science*, **416**, L1085 (1998).
15) T. Kondo, Y. Honma, J. Oh, T. Machida, and J. Nakamura: *Phys. Rev. B*, **82**, 153414 (2010).
16) M. Ziatdinov, S. Fujii, K. Kusakabe, M. Kiguchi, T. Mori, and T. Enoki: *Phys. Rev. B*, **89**, 155405 (2014).
17) O. Yazyev and L. Helm: *Phys.Rev. B*, **75**, 125408 (2007).
18) B. Wang and S. Pantelides: *Phys. Rev. B*, **86**, 165438 (2012).
19) P. Hohenberg and W. Kohn: *Phys. Rev.*, **136**, B864 (1964).
20) W. Kohn and L. Sham: *Phys. Rev.*, **140**, A1133(1965).
21) J. P. Perdew, K. Burke and M. Ernzerhof: *Phys. Rev. Lett.*, **77**, 3865(1996).
22) M. Frisch, G. Trucks, H. Schlegel et al.: Gaussian 03 package software, Gaussian Inc. Wallington CT USA (2009).
23) R. Ditchfield, W. Hehre and J. Pople: *J. Chem. Phys.*, **54**,724 (1971).
24) F. Hund: *Z. Phys.*, **33**, 345 (1923).
25) N. Ota: *J. Magn. Soc. Jpn.*, 38, 107-110 (2014).